\documentstyle[12pt]{article}

\input epsf

\begin{document}
\begin{titlepage}
\today          \hfill
\begin{center}
\hfill    OITS-689 \\

\vskip .05in

{\large \bf 
Improved 
GUT and SUSY breaking by the same field
}
\footnote{This work is supported by DOE Grant DE-FG03-96ER40969.}
\vskip .15in
Kaustubh Agashe \footnote{email: agashe@oregon.uoregon.edu} 
\vskip .1in
{\em
Institute of Theoretical Science, \\
University
of Oregon, \\
Eugene, OR 97403-5203, USA}
\end{center}

\vskip .05in

\begin{abstract}
In a previous paper \cite{gutsusy}, 
we presented 
a 
model in which the {\em same} modulus field breaks 
SUSY
and a simple
GUT
gauge group, 
and 
which has 
{\em dynamical} origins
for {\em both} SUSY breaking and GUT scales. 
In this model, the supergravity (SUGRA)
and gauge mediated contributions to
MSSM scalar and gaugino
masses are comparable -- this enables a realistic
spectrum to be attained since the gauge mediated contribution
to the right-handed (RH) slepton (mass)$^2$ (at the
weak scale) by itself (i.e., neglecting SUGRA
contribution to sfermion {\em and} gaugino masses) is negative. 
But, in general,
the SUGRA
contribution to sfermion masses 
(from non-renormalizable contact K\"ahler terms)
leads to flavor violation. In this paper, we use
the recently proposed idea of gaugino mediated SUSY breaking 
($\tilde{g}$MSB) to improve
the above model. With MSSM matter and SUSY breaking fields 
localized on
separate branes in an extra dimension of size $R \sim 5 M^{-1}_{Pl}$
(in which gauge fields propagate), 
the SUGRA contribution to sfermion masses (which violates
flavor) is suppressed.
As in $4$ dimensions, 
MSSM gauginos acquire {\em non}-universal masses from both SUGRA
and gauge mediation -- gaugino masses (in particular
the SUGRA contribution to gaugino masses), in turn, generate acceptable
sfermion masses through 
renormalization group evolution; the phenomenology is discussed
briefly. We also point out that a)
in
models where SUSY is broken by a GUT non-singlet field, there is,
in general,
a contribution to MSSM gaugino (and scalar)
masses from the coupling to heavy gauge multiplet
which {\em might} be {\em comparable} to the SUGRA contribution and
b) 
models of gauge mediation
proposed earlier which also have negative RH slepton (mass)$^2$
can be rendered viable using the $\tilde{g}$MSB idea.
\end{abstract}

\begin{center}


\end{center}

\end{titlepage}

\newpage
\renewcommand{\thepage}{\arabic{page}}
\setcounter{page}{1}

\section{Introduction}
Two central
issues in supersymmetric grand unified theories (SUSY-GUT's)
are a) mechanism of
SUSY breaking (including the origin of the SUSY
breaking scale) and mediation of SUSY breaking to
the SM superpartners and b) mechanism of GUT symmetry breaking
(down to the SM gauge group)
and the origin of the GUT scale (denoted by
$M_{GUT}$) $\sim 2 \times 10^{16}$ GeV (the
energy scale at which SM gauge couplings unify with the MSSM particle 
content).

In \cite{gutsusy}, we presented a model in which both these
symmetries (SUSY and a {\em simple}
GUT gauge group) 
are broken by the same modulus field
(i.e., by the {\em same} scalar potential). 
A non-zero vev for the $F$-component of this field is generated 
{\em dynamically}
breaking SUSY.
The GUT scale which is the vev of the scalar 
($A$-)component of the {\em same}
field (at the minimum of the potential)
is determined ({\em
dynamically}) 
by an ``inverted hierarchy''
mechanism.
Therefore the GUT scale is {\em naturally}
both larger than the SUSY breaking scale (which is required
for the vev of the scalar component to be {\em calculable}
in perturbation theory)
{\em and} smaller than the Planck scale.
This is the first example of its kind in the
literature. \footnote{In \cite{drees,hirayama} also,
SUSY and a GUT gauge group are broken by the same field.
However, in \cite{drees},
SUSY breaking is {\em not} dynamical so that
a very small superpotential coupling
is required to explain 
the smallness of the SUSY breaking scale compared to $M_{Pl}$ and also 
if all superpotential couplings are of the same order, then 
$M_{GUT} \sim M_{Pl}$. 
In \cite{hirayama}, the GUT gauge
group
is {\em not} simple so that gauge coupling unification is {\em not}
a prediction of the model and also an assumption about a ({\em
non-calculable})
K\"ahler potential is required for the model to work.} 

In this model, there are two comparable
contributions to the MSSM scalar and gaugino masses -- one is mediated by
supergravity (SUGRA) and the other by gauge interactions. This 
is crucial to achieving a realistic sfermion 
mass spectrum since if we neglect 
the SUGRA contribution, then the
(gauge mediated contribution to) the right-handed (RH) slepton
(mass)$^2$ (at the weak scale) is negative. However, the SUGRA contribution
to sfermion (mass)$^2$ is, in general, arbitrary in flavor space
giving unacceptable rates for flavor
changing neutral currents (FCNC's).
We will argue that, in general, this ``problem''
will be present in {\em any} model where SUSY and a GUT symmetry
are broken by the same field.

In this paper, we show how the above model can be ``improved'', i.e.,
how ``flavor-conserving'' and 
positive sfermion (mass)$^2$
can be attained using the recently proposed idea of 
gaugino mediated SUSY breaking \cite{gmsb1, gmsb2}. 

\section{Review of model and the problem}
\subsection{Model}
\label{review}
We begin with a brief review of the model of \cite{gutsusy}.
The gauge group is:
\begin{equation}
SU(6)_{GUT} \times SU(6)_S
\end{equation}
and 
the field content is listed in Table \ref{content}.

\renewcommand{\arraystretch}{1}
\begin{table}
\begin{center}
\begin{tabular}{c|c|c}
 & $SU(6)_{GUT}$ & $SU(6)_S$ \\ \hline
$\Sigma$ & ${\bf 35}$ & ${\bf 1}$ \\
$Q$, $\bar{Q}$ & ${\bf 6}$, $\bar{{\bf 6}}$ & ${\bf 6}$, $\bar{{\bf 6}}$ \\
$H_{1,2}$, $\bar{H}_{1,2}$ & ${\bf 6}$, $\bar{{\bf 6}}$ & ${\bf 1}$ \\
$h_{1,2}$, $\bar{h}_{1,2}$ & ${\bf 6}$, $\bar{{\bf 6}}$ & ${\bf 1}$ \\
$N_i$ & ${\bf 15}$ & ${\bf 1}$ \\
$\bar{P}_{1,2i}$ & $\bar{{\bf 5}}$ & ${\bf 1}$ \\
$Y$, $\bar{Y}$ & ${\bf 15}$, $\bar{{\bf 15}}$ & ${\bf 1}$ \\
$S_{1,2,3}$, $X_{1,2}$, $\bar{X}_{1,2}$ & ${\bf 1}$ & ${\bf 1}$ \\ 
\end{tabular}
\end{center}
\caption{The particle content of the model. $i=1,2,3$ is a generational 
index.}
\label{content}
\end{table}

The core part of the superpotential is:
\begin{equation}
W_1 = \lambda_Q \Sigma Q \bar{Q} + \frac{\lambda_{\Sigma}}{3} \Sigma ^3.
\label{w1}
\end{equation}
Along the flat direction parametrized by tr $\Sigma ^2$, the vev of 
$\Sigma$ is:
\begin{equation}
\langle \Sigma \rangle = \frac{v}{\sqrt{12}} \;
\hbox{diag}[1,1,1,-1,-1,-1]
\label{sigmavev}
\end{equation}
which breaks $SU(6)_{GUT}$ to $SU(3) \times SU(3) \times U(1)$
at the scale $v$.
We identify one unbroken $SU(3)$ with $SU(3)_{color}$ and show later how to
break the other $SU(3) \times U(1)$ to $SU(2)_{weak}
\times U(1)_Y$ at the same scale ($v$). Thus
we identify 
the value of $v$ at the minimum
of the potential (see later) with
the GUT scale (i.e., the scale at which SM gauge couplings meet).
The $\Sigma$ vev
gives mass to $Q$, $\bar{Q}$ so that below the scale $v$,
the only massless fields are the flat direction $v$ 
(we will denote both the chiral superfield and its scalar component
by $v$) and the
pure gauge theory $SU(6)_S$ (the other components of $\Sigma$
either get a mass through the $\Sigma ^3$ term or are eaten
by the broken generators of $SU(6)_{GUT}$; see \cite{gutsusy}
for details).

The pure $SU(6)_S$ gauge theory undergoes gaugino condensation
and generates the superpotential:
\begin{equation}
W = 6 \Lambda_L^3 = \sqrt{3} \lambda_Q \Lambda ^2 v,
\end{equation}
where $\Lambda$ and $\Lambda _L$ are the dynamical scales of 
the high and low energy $SU(6)_S$, respectively (they are related
by one-loop matching at the scale $v$, assuming $v \gg \Lambda$).
Thus, below the scale $\Lambda _L$, the only massless field is $v$ with the
above superpotential so that
SUSY is broken since $F_v \sim \Lambda ^2$.
But, with the canonical (tree-level) K\"ahler potential ($v^{\dagger}
v$), the vev of the scalar ($A$-)component of $v$ 
(i.e., the GUT scale) is undetermined
since the scalar potential is flat $\sim \Lambda ^4$.
The dominant corrections to the K\"ahler potential (and hence
the scalar potential) are given by the
wavefunction renormalization of $\Sigma$ (denoted by $Z$) so that:
\begin{equation}
V (v) \sim \frac{\Lambda^4}{Z(v)}.
\end{equation}
Since $v \gg \Lambda$, we can compute $Z(v)$ using perturbation theory.
In renormalization
group (RG) evolution, at one-loop, 
$Z(v)$ receives contributions from the Yukawa coupling(s)
($\lambda _{\Sigma,Q}$)
and the $SU(6)_{GUT}$ gauge coupling -- the former (latter) tends
to decrease (increase) $Z(v)$ as $v$ increases. Thus,
if the gauge coupling dominates at small $v$ 
whereas the Yukawa coupling is larger at high scales (as is natural if
$SU(6)_{GUT}$ is asymptotically free), then
the potential can develop a minimum. Furthermore,
the minimum can be (naturally) at a value of $v 
\gg \Lambda$ since
$Z$ and both couplings depend logarithmically on $v$: at the minimum,
we require that 
$v \sim 10^{16}$ GeV (to obtain the correct GUT scale) whereas
$\Lambda \sim 10^{10} - 10^{11}$ GeV so that MSSM sparticles
have masses $\sim 100 \; \hbox{GeV} - 1$ TeV (see later).
\footnote{This is only a local minimum since there is a supersymmetric
minimum with $\langle \Sigma \rangle \sim \langle Q
\rangle \sim \langle \bar{Q} \rangle
\sim \Lambda$ \cite{gutsusy}. However since, at the local minimum
$v \gg \Lambda$ the tunneling rate
from the local minimum to the ``true'' vacuum
is very small.} 
Thus,
this inverted
hierarchy mechanism \cite{witten}
can generate a GUT scale, $M_{GUT}$, 
much larger
than the SUSY breaking scale, $\Lambda$ (which is also required for
the perturbative 
calculation mentioned above to be valid)
\footnote{It should be possible to contruct models based
on (say) $SO(10)$ along similar lines.}. We can also view this
as ``generating'' 
the GUT scale from the Planck scale ($M_{Pl} \sim 2 \times 10^{18}$
GeV)
as follows. We can choose
$Z = 1$ (canonical normalization) at $M_{Pl}$ and RG evolve $Z$ to lower
energies. Since $Z$ and the gauge and Yukawa couplings vary logarithmically
(``slowly'') with energy, 
(we can choose $O(1)$ couplings at $M_{Pl}$
such that) 
$Z$ reaches a maximum at an energy scale $\sim M_{GUT}$ which is
``much'' smaller (i.e., by two orders of magnitude)
than $M_{Pl}$ \cite{gutsusy}.

To break the other $SU(3) \times U(1)$ to $SU(2)_{weak} \times U(1)_Y$
(and to get the usual light Higgs doublets)
we add:
\begin{equation}
W_2 = \sum_{i=1}^{2} S_i (H_i \bar{H}_i - \Sigma^2),
\end{equation}
\begin{equation}
W_3 = \sum_{i=1}^{2}
H_i (\Sigma + X_i) \bar{h}_i + \sum_{i=1}^{2} \bar{H}_i
(\Sigma + \bar{X}_i) {h}_i,
\end{equation}
\begin{equation}
W_4 = \frac{1}{M} \left( \left( H_1 \bar{H}_1 \right)
\left( H_2 \bar{H}_2 \right) - \left( H_1 \bar{H}_2 \right)
\left( H_2 \bar{H}_1 \right) \right),
\label{higherdim}
\end{equation}
\begin{equation}
W_5 = S_3 \left( H_1 \bar{H}_2 - H_2 \bar{H}_1
\right).
\end{equation}

The role of these $W$'s is as follows (for details, see \cite{gutsusy}).
$W_2$ forces $H$, $\bar{H}$ to acquire vev's. With 
$\langle H \rangle = \langle \bar{H} \rangle \sim v \; (1,0,0,0,0,0)$,
$SU(3) \times U(1)$ is broken to $SU(2) \times U(1)$.
$W_3$ forces $X$, $\bar{X}$ to acquire vev's such that
only the triplets in $H$'s get a mass with those in $\bar{h}$'s
(this is the ``sliding singlet'' mechanism). \footnote{Any symmetries
which allow the
terms $\Sigma ^3$, $S \Sigma^2$ and $S H \bar{H}$
(which we need to obtain the desired vev's), 
also allow the term $\Sigma
H \bar{H}$ which
spoils the above pattern of vev's; in addition, there might be higher
dimension operators (allowed by the same symmetries) 
which might spoil the sliding singlet mechanism and/or the pattern of vev's.
Thus, this model is only ``technically'' natural, i.e., the superpotential
is {\em not} the most general one allowed by symmetries.}
A pair of doublets in $H$, $\bar{H}$ is eaten in the gauge symmetry
breaking while a pair of doublets in $h$, $\bar{h}$
gets a mass with doublets in $\Sigma$ (due to the $H$, $\bar{H}$ vev).
This leaves
two pairs of doublets massless: the one in $H$, $\bar{H}$ acquires
mass through $W_4$ and the one in $h$, $\bar{h}$ is the usual pair
of Higgs doublets. $W_5$ gives a required
constraint between the vev's of $H_1$,
$\bar{H}_1$ and $H_2$,
$\bar{H}_2$.

To complete the model, the SM quarks and leptons are obtained through:
\begin{eqnarray}
W_{6} & = & N_i (\bar{P}_{1j} \bar{H}_1 + \bar{P}_{2j} \bar{H}_2) +
N_i (\bar{P}_{1j} \bar{h}_1 + \bar{P}_{2j} \bar{h}_2) \nonumber \\
 & & + N_i N_j Y + (X_1 + X_2) Y \bar{Y} + \bar{Y} (H_1 h_1 - H_2 h_2).
\label{w6}
\end{eqnarray}
For each generation, 
the vev's of $\bar{H}$'s gives a mass to
one combination of $\bar{{\bf 5}}$ (under $SU(5)$) in 
$\bar{P}_{1,2}$ with the ${\bf 5}$ 
(under $SU(5)$) of $N$ leaving one $\bar{{\bf 5}}$
in $\bar{P}_{1,2}$ and ${\bf 10}$ in $N$ massless -- these
are the quarks and leptons. The other terms in $W_6$ generate their
Yukawa couplings.

\subsection{Problem with sfermion spectrum}
\label{problem}
The MSSM scalars and gauginos
acquire masses $\sim \left[ \alpha_{GUT} / \pi  
\right] \left[ F_v / M_{GUT} \right] \sim 10^{-2} \left[ F_v / M_{GUT} 
\right]$ 
through gauge mediation (GM) by coupling at one or two-loops to
two sources \footnote{The ``loop'' factor for these masses is
$\sim
\alpha _{GUT} / (4 \pi)$, but there is usually an enhancement
from group theory and/or
large number of messengers which effectively makes
the loop factor
$\sim
\alpha _{GUT} / \pi \sim 10^{-2}$ (for $\alpha _{GUT} \approx 0.04$) --
this estimate suffices for comparison to SUGRA mediated masses (see later).
The
precise expressions for the gauge mediated masses
are given in Eqs. (\ref{gmgaugino}) and
(\ref{gmscalar}).
}: one
is the ``matter'' messengers -- the
$Q$, $\bar{Q}$ fields
and the heavy components of $H$, $\bar{H}$, $h$ and $\bar{h}$ fields 
which as usual have a SUSY breaking mass spectrum due to the coupling
to $v$. \footnote{With the addition of $W_{2,3}$, the flat direction $v$
is a combination of $\Sigma$, $H$, $\bar{H}$, $X$ and $\bar{X}$.}
The other source, usually
referred to as ``gauge'' messengers, 
is the heavy part of the $SU(6)_{GUT}$ gauge multiplet
which also has a SUSY breaking spectrum since the field(s) breaking
the GUT gauge group ($\Sigma$, $H$ and $\bar{H}$) have a non-zero
$F$-component.

Using the technique of
\cite{gr}, the MSSM gaugino masses are given by
($A$ denotes the gauge group):
\begin{equation}
M_A (\mu) \approx \frac{\alpha_A (\mu)}{4 \pi} \frac{F_v}{v} \left(b_A - b_6
\right),
\label{gmgaugino}
\end{equation}
where $b_A$'s are the beta functions of the SM gauge groups below
the GUT
scale and $b_6$ is the beta function of the $SU(6)_{GUT}$ above the
GUT scale. The gaugino masses are {\em non}-universal since the messengers
(both gauge and matter)
are {\em not} in complete GUT multiplets.

The matter messengers give a positive contribution to the scalar (mass)$^2$
as usual, but the gauge messenger contribution 
is typically negative (for
all scalars) so that
most scalars have negative (mass)$^2$
at the GUT scale. Of course, in RG scaling to the weak scale, the
sfermion (mass)$^2$ get a positive contribution from the gaugino masses.
The gauge mediated MSSM sfermion (other than stop) 
(mass)$^2$ at the scale $\mu$ are given by (again
using the technique of \cite{gr}):
\begin{eqnarray}
m^2_i (\mu)& \approx & \frac{1}{16 \pi^2} \left( \frac{F_v}{v} \right)^2
\times \nonumber \\
 & & \left( \sum_A \frac{2 C^i_A}{b_A} \left(
\alpha_A^2(\mu) \left( b_6 - b_A \right) ^2 - b_6^2 \alpha_6^2 \right)
+ 2 C^i_6 b_6 \alpha_6^2 \right),
\label{gmscalar}
\end{eqnarray}
where $C^i_A$ is the quadratic Casimir invariant
for the scalar $i$ under the gauge
group $A$, {\it i.e.}, $4/3$, $3/4$ for fundamentals of
$SU(3)_c$, $SU(2)_L$  respectively and $3/5 \;
Y^2$ for
$U(1)_Y$. $C^i_6 = 35/12$ for fields in ${\bf \bar{5}}$ of
$SU(5)$ (${\bf \bar{6}}$ of
$SU(6)_{GUT}$) and $14/3$ for fields in ${\bf 10}$ of
$SU(5)$ (${\bf 15}$ of
$SU(6)_{GUT}$).
The beta function for $SU(N_c)$ group is defined as $3 N_c - N_{f,eff}$,
where
$N_{eff}$ is the ``effective'' number of flavors.
$\alpha_6$ is the $SU(6)$
coupling at the GUT scale. With the particle content in Table \ref{content},
we get $b_6 = -11$.

In this model, it turns out that the gauge
mediated contribution to RH slepton (mass)$^2$, Eq. (\ref{gmscalar}), is
negative at the weak scale, i.e., the
positive bino mass contribution (which is {\em not} an
independent parameter) is not enough
to make the RH slepton (mass)$^2$ positive
while all other sfermion (mass)$^2$ are 
positive due to the larger wino/gluino mass contribution.

So far, the SUGRA contribution to MSSM scalar and gaugino masses
has been neglected.
The SUGRA contribution to MSSM sfermion (mass)$^2$ from the operators:
\begin{equation}
{\cal L} \sim \int d^4 \theta 
c_1 \; \frac{ X^{\dagger} X \Phi _i ^{\dagger} \Phi _j } 
{M_{Pl}^2},
\label{sugrascalar}
\end{equation}
where $X$ is any SUSY breaking field ($\Sigma$, $H_i$, $X_i$ etc.)
and $\Phi$ is a MSSM matter field,
is $\sim \left[ F_v / M_{Pl} \right] ^2 \sim 10^{-4} \left[ F_v / M_{GUT}
\right] ^2$ (
assuming $c_1 \sim O(1)$ and
$M_{Pl} \sim 2 \times 10^{18}$ GeV). 
Thus, the SUGRA and gauge mediated contributions (to sfermion masses) are 
comparable
so that the combined RH
slepton (mass)$^2$ can still be positive
(provided the SUGRA contribution is positive and a bit larger
than the GM contribution) and a realistic
spectrum can be achieved. However, the SUGRA contribution
(Eq. (\ref{sugrascalar}))
violates flavor -- in general,
the off-diagonal terms in the
sfermion (mass)$^2$ matrices (in flavor space) will be $O(
\left[ F_v / M_{Pl} \right] ^2)$
which clearly result in too large SUSY contributions
to FCNC's. There is also a SUGRA contribution to 
MSSM gaugino 
masses ($\sim F_v / M_{Pl}$) comparable
to the GM contribution (see ($4D$ equivalent of) 
Eq. (\ref{sugragaugino})) which, in turn, contributes to scalar masses
in RG scaling to the weak scale; this contribution to sfermion (mass)$^2$
is positive and flavor-conserving.

From the above discussion, it is clear that generic models 
(i.e., not just the one in \cite{gutsusy})
in which
the same field breaks SUSY and a GUT symmetry will have the same 
problem -- at the scale $M_{GUT}$, there will be a contribution
(which is typically negative) 
to the scalar (mass)$^2$ from the
heavy gauge multiplet
so that, at least for the RH slepton,
the gauge mediated (mass)$^2$
at the weak scale 
might be negative. Also, even if
the gauge mediation contribution is positive,
the SUGRA contribution (Eq. (\ref{sugrascalar})) to the
sfermion (mass)$^2$ is comparable 
which leads to flavor violation, in general.
It is clear that the latter is a problem also 
in models of GM with messenger scale
close to the GUT scale (having nothing to do with GUT symmetry 
breaking), for example the model of \cite{pt}.

\section{Improved model using an extra dimension}
We now show how
to improve the above model, i.e., how to obtain
positive and (at the same time) flavor-conserving sfermion (mass)$^2$
using the framework of gaugino mediated SUSY
breaking ($\tilde{g}$MSB) \cite{gmsb1, gmsb2}.

Consider the following
embedding of this model in a $5$-dimensional ($5D$)
theory. 
The MSSM matter fields (i.e., $N$ and $\bar{P} _{1,2}$) are 
localized on a ``$3$-brane'' (``MSSM matter'' brane) whereas the 
$SU(6)_{GUT}$ gauge
fields and $Y$, $\bar{Y}$,
$h$, $\bar{h}$ fields propagate in the extra dimension.
SUSY breaking fields, i.e., $\Sigma$, $H$, $\bar{H}$,
$X$ and $\bar{X}$ and $Q$, $\bar{Q}$
are localized on a different $3$-brane
(``SUSY breaking'' brane) which is separated from the MSSM matter brane 
by a distance $R \sim 3 \; M^{-1}$ in the extra dimension,
where $M$ is the ``fundamental'' ($5D$) Planck scale. For
simplicity we also assume that $R$ is the size of the extra dimension.
The $S_{1,2,3}$ fields and the $SU(6)_S$ gauge fields
can either propagate in the bulk or be localized on the 
SUSY breaking brane.

In the effective $5D$ field
theory below $M$ ($\sim 10^{18}$ GeV, see later),
direct couplings between fields on
the matter brane
and SUSY breaking brane
are forbidden since such couplings 
are not local \cite{rs}. 
Of course, such operators might be generated by integrating out
bulk states with mass $\sim M$, but such effects will suppressed
by a Yukawa factor
$\sim e^{-R M }$ due to the (position
space) propagator of the bulk state \cite{rs}.
In particular, the
coefficient of the operator in Eq. (\ref{sugrascalar}) is 
$\stackrel{<}{\sim}10^{-2}$. Thus, the SUGRA contributions to the 
MSSM squark and slepton (mass)$^2$ from
contact K\"ahler terms are suppressed by this factor
relative to other contributions (see later).

The superpotential couplings written in section \ref{review}
are all allowed, except
for the $N \bar{P}_{1,2} \bar{H}_{1,2}$ coupling
in Eq. (\ref{w6}) ($N$, $\bar{P}_{1,2}$
and
$\bar{H}$ fields are localized on different branes). This coupling
in the model of section \ref{review} 
gives an $O(M_{GUT})$ mass
to ${\bf 5}$ (under $SU(5)$) of $N$ with a $\bar{{\bf 5}}$ 
(under $SU(5)$) in $\bar{P}_{1,2}$
while the massless $\bar{{\bf 5}}$ in $\bar{P}_{1,2}$ 
and ${\bf 10}$ in $N$ are
the usual quarks
and lepton fields.
Thus, in the above framework, a $({\bf 5} + \bar{{\bf 5}})$ (for each
generation) is massless in addition to the usual quarks and leptons.
The beta-functions
of SM gauge group ($b_A$'s) and hence the GM contribution to scalar and 
gaugino masses (in the $4D$ theory; see later)
depends on whether these
addtional fields are light or not.
We assume that
the extra $({\bf 5} + \bar{{\bf 5}})$'s
acquire mass $\sim O(M_{GUT})$ by some
mechanism. \footnote{For example, the extra $({\bf 5} + \bar{{\bf 5}})$'s
might couple to additional fields
or another possibility is that
the operator $\int d^2 \theta \left(
N \bar{P} \bar{H} X / M \right)$ is
generated with a coupling $\sim e^{-RM}$ by exchange of bulk
(``string'') states -- this operator will gives a mass
$O( e^{-3} \; M_{GUT} / M ) \sim 10^{12}$ GeV to the extra 
$({\bf 5} + \bar{{\bf 5}})$. In the latter case, there is an additonal
messenger scale for
GM
$\sim 10^{12}$ GeV since the $({\bf 5} + \bar{{\bf 5}})$
in $N$, $\bar{P}_{1,2}$ have a non-supersymmetric spectrum due
to the coupling to $\bar{H}$, $X$ -- in the $4D$ model
of section \ref{review} these messengers were at the scale $M_{GUT}$.}

We require 
the (usual Higgs doublets in)
$h$, $\bar{h}$ fields to couple to matter fields so that
Yukawa couplings can be
generated. These Higgs fields should also couple
to the GUT
symmetry breaking fields (and hence SUSY breaking fields, in this case), 
which are localized on a different brane, 
so that the usual Higgs
doublet-triplet splitting can be achieved. Thus, Higgs fields 
$h$, $\bar{h}$ have to
propagate in the bulk.

The MSSM gauginos get a mass from the following ($5D$) SUGRA
interactions:
\begin{equation}
{\cal L} \sim \int d^2 \theta \left( c_2 \; \frac{\Sigma W _{\alpha} W
^{\alpha}}{M^2} +
c_3 \; \frac{(X + \bar{X}) W _{\alpha} W ^{\alpha}}{M^2} \right) + h.c.,
\label{sugragaugino}
\end{equation}
where $W _{\alpha}$ is a $5D$ gauge field
and $c_{2,3} \sim O(1)$.
The contribution to MSSM gaugino masses
from the singlet is universal
while that from $\Sigma$ is non-universal:
$M_1 \; : \; M_2 \; : \; M_3 = 1  \; : \; 5 \; : \; -5$.
When
these operators and also the operators $\sim \int d^4 \theta \;
X^{\dagger} X / M^5 \; W_{\alpha} D^2 W^{\alpha}$ (which generate
a SUSY breaking gaugino wavefunction) are inserted in one-loop diagrams,
we get 
contributions 
to sfermion (mass)$^2$ $\sim g_{5D}^2 / ( 4 \pi^2 )
\left[ F_v / M^2 \right] ^2 1 / R^3$ and $\sim g_{(5D)}^2 / ( 4 \pi^2 ) \;
F_v^2 / M^5 \; 1 / R^4$, respectively
\cite{gmsb1, gmsb2} ($g_{(5D)}$ is the
$5D$ gauge coupling)
which are {\em finite}
due to the spatial separation
of the SUSY breaking and the MSSM matter branes in the
$5D$ theory. \footnote{In other words,
these contributions to sfermion (mass)$^2$
are the effect of integrating out the extra dimension,
i.e., the Kaluza-Klein excitations of the gauge fields.
}
Using Eqs. (\ref{Plmatching}) 
and (\ref{gmatching}) (see below), these one-loop contributions 
to sfermion
(mass)$^2$ are of order $\alpha _{(4D)} / \pi \left[ F_v / M_{Pl} \right] ^2
\; 1/ (MR)$ and $\alpha _{(4D)} / \pi \left[ F_v / M_{Pl} \right] ^2
\; 1/ (MR)^2$, respectively and thus are negligible compared
to GM contribution at $M_{GUT}$ and gaugino mass contribution
in RG scaling to the weak scale (see below).

When the extra dimension
is compactified (i.e., the
Kaluza-Klein (KK) excitations of supergravity and other bulk states
are integrated out), we get an effective $4D$ field theory below
the scale $R^{-1} \sim 1/3 M$.
The $4D$ and $5D$ Planck scales are related by
\begin{equation}
M_{Pl}^2 \sim M^3 R \sim 3 \; M^2 
\label{Plmatching}
\end{equation}
so that $M \sim M_{Pl} / \sqrt{3} \sim 10^{18}$ GeV and $R^{-1}
\sim M_{Pl} /5 \sim 4 \times 10^{17}$ GeV.
\footnote{For simplicity,
we assume that any other extra dimensions have size $M^{-1}$ so that
$M$ is the {\em fundamental} quantum gravity (``string'') scale and
the inverted hierarchy mechanism generates the hierarchy between
$M$ and $M_{GUT}$ (\`a la the
hierarchy between $M_{Pl}$ and $M_{GUT}$ in $4D$).
It is easy
to extend this framework to one with {\em more than one} extra
dimensions of size {\em slightly larger} 
than the fundamental length scale. However,
in that case, the fundamental (say, $(4 + n) D$) Planck scale,
$M$, might be smaller than $10^{18}$ GeV 
(since $M_{Pl}^2 \sim M^{n+2} R^n$ for $n$ extra dimensions of size $R$)
so that
the motivation for the inverted hierarchy mechanism (to explain
$M_{GUT} / M$) is a bit weaker.
}
{\it In what follows, we assume} $M \sim M_{Pl}$.
It was shown in \cite{luty} that 
the exchange of supergravity Kaluza-Klein (KK) excitations 
also does not lead to contact K\"ahler terms of order $1 / 
M _{Pl} ^2$ (of the form
Eq.
(\ref{sugrascalar})).
The $4D$ and $5D$ gauge couplings are related by
\begin{equation}
g_{(4D)}^2 \sim \frac{g_{(5D)}^2}{R}.
\label{gmatching}
\end{equation}
The $4D$ (zero-mode) 
gaugino mass is given by $ \sim c_{2,3} \;
F_v / M^2 \; 1 / R 
\sim c_{2,3}
\; F_v / ( \sqrt{3} \; M_{Pl} )$, i.e., $O ( F_v / M_{Pl} )$ \cite{gmsb1}.

The anomaly mediated contribution to MSSM scalar
and gaugino masses \cite{rs, glmr} is \\
$\sim F _v / M_{Pl} \; \alpha / \pi$ and thus can be neglected
in comparison to 
SUGRA contribution to gaugino
masses,  
GM contributions (at $M_{GUT}$) to gaugino
and scalar masses and gaugino mass
contribution to scalar masses in RG scaling to the weak scale
(see below).
There is also a contribution to scalar
(mass)$^2$ at one-loop $\sim
1/ \left( 16 \pi^2 \right) F_v^2 / (R^2 \; M_{Pl}^4)$ from integrating
out SUGRA KK modes \cite{rs}; for $MR \sim 3$, this is comparable
to the anomaly mediated contribution and thus can be neglected. \footnote{
Both these contributions to sfermion masses and also the one-loop
gaugino contribution in the $5D$ theory mentioned above
are flavor-conserving.}

At the scale $M_{GUT} \sim 10^{16}$ GeV
\footnote{As before, the GUT scale is determined
by the one-loop corrections to the wavefunction ($Z$)
of $\Sigma$. Between
the energy scales $M$ and $R^{-1}$, the $SU(6)_{GUT}$
gauge coupling (rather $\alpha _6 ^{-1}$) (and similarly 
$Z$, $\lambda _{Q, \Sigma}$)
``runs'' with a power of energy 
since the gauge theory
is $5D$ whereas below $R^{-1} \sim 4 \times 10^{17}$
GeV, we have the usual ($4D$) RG scaling. 
In any case,
the inverted hierarchy mechanism can result in a minimum
of the potential at $v \sim 10^{16}$ GeV $\ll M$;
most of the RG scaling 
of $Z$, $g_6$ and $\lambda _{\Sigma,Q}$ 
from $M$ to $\sim 10^{16}$ GeV
is the usual $4D$ evolution.}, the heavy gauge multiplet
and the matter messengers are integrated out (as in
section \ref{problem}) generating
the contributions to the scalar (mass)$^2$ (at two-loops) and gaugino masses
(at one-loop), Eqs. 
(\ref{gmgaugino}) and (\ref{gmscalar}). 
\footnote{
In computing these ``threshold''
corrections, the SUGRA contribution $\sim
F_v / M_{Pl}$ 
to the SUSY breaking masses 
of the heavy gauge 
multiplet and the matter messengers ($Q$, $\bar{Q}$ etc.)
can be neglected in comparison to the contribution $\sim F_v / M_{GUT}$
from direct coupling to $v$.
}
There are also
contributions to sfermion and gaugino masses
obtained by replacing the zero-mode ($4D$) gauge fields in the above loop
diagrams by
KK gauge fields which have masses $\sim k R^{-1}$.
The one-loop diagrams with KK gauge fields and gauge messengers
give (zero-mode) gaugino masses
$\sim \sum _k \; \alpha / \pi \; F_{v} v / \left( k R^{-1} \right) ^2$
(since the $R$-symmetry breaking scale is $v$) and the two-loop
diagrams with KK gauge fields 
(and either matter or gauge messengers)
give sfermion (mass)$^2 \sim \sum _k \; \left( 
\alpha / \pi \right)^2 \; F_v^2 / \left( k R^{-1} \right) ^2$. 
\footnote{Strictly speaking, since this is the effect of integrating
out the KK gauge fields, these contributions to sfermion and
gaugino masses appear at/above the scale $\sim R^{-1}$.}
Since $\sum _k
\;
1/ k^2$ is convergent for the case of one extra dimension, we see that 
these contributions are  
suppressed by a factor $\sim 1/ \left(
M_{GUT} R \right)^2 \sim O(100)$ compared to Eqs.
(\ref{gmgaugino}) and (\ref{gmscalar}).
As mentioned earlier,
GM contributions to gaugino
and scalar masses (at $M_{GUT}$) (Eqs.
(\ref{gmgaugino}) and (\ref{gmscalar})) and the SUGRA contribution
to gaugino mass are all of order $F_v / M_{Pl}$.
In RG scaling to the weak scale, the gaugino masses give an
additional (positive) 
contribution to the
sfermion (mass)$^2$ $\sim \alpha / \pi \;
\ln \left( M_{GUT} / m_Z \right) \; \left[
F_v / M_{Pl} \right] ^2 \sim \left[
F_v / M_{Pl} \right] ^2$.

It is clear that if $R \sim 3 M^{-1}$, then the SUGRA 
contribution to the sfermion (mass)$^2$ at the high scale 
from contact K\"ahler terms
(Eq. (\ref{sugrascalar}) with a coefficient
$\sim e^{-MR}$) can be neglected in
comparison to the GM contribution at $M_{GUT}$
and the contribution generated by gaugino masses in RG scaling. 
This
means that sfermion masses
conserve flavor, i.e., sfermion with the same gauge quantum numbers
are degenerate at the 
$O( e^{-3}) \sim $ percent level. \footnote{This 
degeneracy is sufficient to evade limits
from $CP$-{\em conserving} flavor-violating processes, for example,
$\mu \rightarrow e \gamma$, $\Delta m_K$ etc. But, 
if the SUGRA mediated sfermion
(mass)$^2$ in Eq. (\ref{sugrascalar}) have 
$O(1)$ phases, 
then 
we need $R \sim 6 \; M^{-1}$ to obtain degeneracy at the $\sim 0.1 \; \%$ 
level so that 
SUSY contributions to
$CP$ {\em and} flavor-violating processes
are suppressed.}

As mentioned earlier, if the SUGRA contribution to sfermion {\em and} gaugino
masses is neglected, then the (gauge mediated contribution to)
RH slepton (mass)$^2$ at the weak scale is negative.
However, in this ($5D$) framework, while
the SUGRA contribution to RH slepton mass at the high scale (Eq.
(\ref{sugrascalar}))
is suppressed, we have to include
the SUGRA contribution ($\sim
F_v / M_{Pl}$) to the bino mass which, in turn, generates
in RG scaling to the weak scale
an {\em additional} (compared to the pure GM case)
positive contribution $\sim \left[
F_v / M_{Pl} \right] ^2$ to the RH slepton (mass)$^2$. This  
can result in a positive
RH slepton (mass)$^2$ at the weak scale,
i.e., unlike
the pure GM case, due to the SUGRA contribution
to the bino mass, the bino mass contribution is
(effectively) {\em independent} of
(and of the same order as) the GM
contribution to the RH slepton mass. \footnote{Thus, the  
role
of the extra dimension here (as in 
\cite{gmsb1, gmsb2}) is to suppress the (flavor-violating)
SUGRA contribution to sfermion masses while allowing SUGRA
contribution to gaugino masses (in particular, in this case, 
bino mass). It is clear that
any other framework which provides these boundary conditions  
also suffices.}

It is obvious
that the same framework (extra dimension of
size $R \sim 3 M^{-1}$) can be used
to suppress (flavor-violating)
SUGRA contribution to sfermion masses 
(Eq. (\ref{sugrascalar})) in {\em any} model
where the GM and SUGRA contributions are comparable, for example,
a model of GM with messenger scale close to the GUT scale 
(where the GM contribution to sfermion (mass)$^2$ is, say, {\em positive})
\cite{pt}. 
As mentioned earlier, {\em any} model
with GUT and SUSY breaking by the same field is likely
to have negative
GM contribution to sfermion (say, RH slepton)
(mass)$^2$; it is clear that (in addition
to suppressing the SUGRA contribution to sfermion masses)
the above idea (the SUGRA contribution to bino mass)
can
be used to obtain positive RH slepton (mass)$^2$.

A slightly modified version of the above model is obtained by
gauging only the $SU(5)$
subgroup of the $SU(6)_{GUT}$ symmetry \cite{gutsusy}.
The superpotential is given by 
$W_1 + W^{\prime}$ with
\begin{equation}
W^{\prime} = h \left( \lambda _1 \Sigma _1 + 
\lambda _{24} \Sigma _{24} \right) \Sigma _{\bar{5}} + \bar{h} \left( 
\bar{\lambda} _1 \Sigma _1 + 
\bar{\lambda} _{24} \Sigma _{24}\right)
\Sigma _5,
\label{wprime}
\end{equation}
where $\Sigma _r$'s  
refer to components of $\Sigma$ transforming 
as 
${\bf r}$ under $SU(5) _{local}$
and $h$, $\bar{h}$ form a $({\bf 5} + \bar{{\bf 5}})$ of $SU(5) _{local}$.
$\Sigma$ and $Q$, $\bar{Q}$ are as usual localized on the 
SUSY breaking brane with $h$, $\bar{h}$ fields in the bulk. 
In the absence of
$W^{\prime}$, $\Sigma$ has a pair of massless color triplets
which are Nambu-Goldstone fields since the full $SU(6)_{GUT}$ is not gauged.
$W^{\prime}$ gives masses to these triplets with those in $h$, $\bar{h}$.
Since $\langle \Sigma _1 \rangle \sim \; \hbox{diag} \; [1,1,1,1,1
]$ whereas $\langle \Sigma _{24} \rangle \sim \; \hbox{diag} \; [
3,3,-2,-2,-2 ]$ in $SU(5)$ space, we can (fine-)tune the
$\lambda _{1,24}$ couplings so that the weak doublets in
$h$, $\bar{h}$ are massless; these will be the usual Higgs doublets.

In this version of the model, doublet-triplet splitting (although
``technically natural'')
is fine-tuned. 
Also, there is a global $SU(6)$ symmetry on the 
SUSY breaking brane, i.e., $W_1$
is $SU(6)$ symmetric, but the couplings of the bulk fields,
$h$, $\bar{h}$ to $\Sigma$ (Eq. (\ref{wprime}))
break this to $SU(5)_{local}$.
The nice feature
compared to the earlier model is
that
quarks and leptons are contained in
the usual $(\bar{{\bf 5}} + {\bf 10})$ of $SU(5)$ (localized
on the matter brane) and so, unlike the gauged 
$SU(6)$ model, ``splitting'' of matter superfields is not required. 
As mentioned earlier, in the gauged $SU(6)$ model, quarks and leptons
are contained in $({\bf 15} + \bar{{\bf 6}} + \bar{{\bf 6}})$ of $SU(6)$
and a $(\bar{{\bf 5}} + {\bf 5})$ (under $SU(5)$) are
made heavy by coupling to $\bar{H}$ in the $4D$ model -- 
this coupling is not allowed
(at the renormalizable level) in the
$5D$ model since $\bar{H}$ and matter fields are localized
on different branes.
In the $SU(5)$ model, there are SUGRA contributions to MSSM
gaugino masses
from both singlet $(\Sigma _1)$ and adjoint ($\Sigma _{24}$) SUSY
breaking fields (in addition to the GM contribution).
Thus, as in the gauged $SU(6)$ model, the MSSM gaugino masses
($M_1$, $M_2$ and $M_3$) are free parameters (see section \ref{pheno}).

\paragraph{Comments on other models:}
Some of the models of gauge mediation in the
literature (\cite{hitoshi, dimopoulos})
also have gauge messengers and
hence negative MSSM scalar (mass)$^2$ at the messenger scale, $M_{mess}$
(the SUGRA contribution is much smaller than the GM
contribution if $M_{mess} \stackrel{<}{\sim} 10^{15}$ GeV).
As usual, the gaugino masses give a positive
contribution in RG scaling to the weak scale; however, at least
RH sleptons still have negative (mass)$^2$ \cite{gr}. The framework 
of $\tilde{g}$MSB 
can also be used to ``resurrect'' these models as follows. \footnote{
This was partly hinted in \cite{gmsb1}.} Suppose the
SUSY breaking fields are
localized 
on a brane different than the MSSM matter brane in an extra dimension
with a compactification scale ($R^{-1}$)
slightly smaller
than $M_{mess}$ (with gauge fields 
in the bulk). \footnote{As before, we assume that the two branes
are maximally separated in the extra dimension.} Then the two-loop
(negative)
gauge mediated contribution
to sfermion (mass)$^2$ at $M_{mess}$ is {\em suppressed} (in
the $5D$ theory) by a factor 
$1 / (RM_{mess})^2 \sim 1/25$ (if $R \sim 5 \; M _{mess} ^{-1}$)
relative
to the $4D$ result, i.e., it is
$\sim 1/25 \; \left[ \alpha / \pi \right]^2
\left[ F / M_{mess} \right] ^2$ \cite{peskin, gmsb1}.
The gaugino masses are the same as in $4D$ \cite{gmsb1}
($\sim \alpha / \pi \;
F / M_{mess}$) and,  
in turn, give a positive $O( \alpha / \pi \;
F / M_{mess} )^2$ contribution to sfermion
(mass)$^2$ in RG scaling to low scales (provided $M_{mess} \gg m_Z$, i.e,
RG logarithm is large enough). Since the (negative)
sfermion (mass)$^2$ at $M_{mess}$ is small compared to this 
RG contribution, the sfermion (including
RH slepton) (mass)$^2$ at the weak scale can be positive.

Above the scale $M_{mess}$, the
theory is $5D$ with MSSM matter fields on a brane 
and gauge fields in bulk. If Higgs fields are
also in the bulk, then in this framework gauge coupling unification
is (approximately)
preserved (with lower unification scale) \cite{ddg}.

We can use a
similar idea to suppress (negative)
gauge mediated sfermion (mass)$^2$ (at $M_{GUT}$)
in the GUT model, i.e,
we can invoke
an extra dimension slightly larger than (inverse of) the GUT scale 
(which is $M_{mess}$ in this case). This will suppress {\em both} the SUGRA
and (negative)
GM contributions to sfermion (mass)$^2$ at the high scale.
But, with (say) $R \sim 5 \; M ^{-1} _{GUT} \sim (4 \times 
10^{15} \; \hbox{GeV})^{-1}$, 
we get the $5D$ Planck scale $M \sim 10 \; M_{GUT}$ 
(using $M_{Pl}^2 \sim M^3 R$)
so that the motivation
for the inverted hierarchy mechanism (to generate $M_{GUT}$ smaller than the
fundamental Planck scale by a factor of $\sim 100$ as before) 
is a bit weaker.
\footnote{An even larger extra dimension (in which
{\em only} gravity propagates) can be used to lower $M$
all the way to $M _{GUT}$ \cite{horava} so that there is {\em no}
hierarchy between the fundamental Planck scale and GUT scale --
of course, in this case 
(as in the case with 
$R \sim 5 \; M ^{-1} _{GUT}$) 
one has to explain the ``hiererchy'' between 
$R^{-1}$ and $M$. 
Here (as in \cite{gutsusy}), instead
we would like to ``explain''  
$M_{GUT} \sim 10^{-2} M$ using the inverted hierarchy
mechanism.}
In addition a new ``hierarchy'',
$R \sim 50 \; M^{-1}$, would have to be explained. 
\footnote{In this model, the RG scaling
of SM
gauge couplings above the energy scale $R^{-1} 
\sim 4 \times
10^{15}$ GeV is $5D$ --
unification of SM gauge couplings
still occurs (Higgs fields also propagate in 
$5D$) but at a scale lower than
the usual $M_{GUT}$ by a factor of $\sim 2$ \cite{ddg}.
For illustrative purposes,
we
assumed above that the unification scale
is still $M_{GUT} \sim 2 \times 10^{16}$ GeV.
In general, we can choose the compactification scale, $R^{-1}$, 
to be much smaller than $10^{15}$ GeV so that unification 
occurs at 
a scale,  
$M^{\prime} _{GUT}$, 
{\em which
depends on} $R^{-1}$ and 
which is much smaller than 
the usual $M_{GUT} \sim 10^{16}$ GeV \cite{ddg}.
However, 
in this case,
one has to ``explain'' why the compactification scale, $R^{-1}$,
is correlated with the GUT scale
($M^{\prime} _{GUT}$) which (in the context of the model
in this paper), in turn, is determined
by a modulus field (i.e, which is {\em not} a
fundamental scale).
Also, if $R \gg M^{-1}$, then the $5D$ gauge couplings (assuming that
$4D$ gauge couplings are $O(1)$) might become non-perturbative (larger
than their strong coupling value) (see Eq. (\ref{gmatching})).
Of course, one faces a similar issue(s) in
trying to ``save'' the GM models above except that in that case
the correlation between $M_{mess}$ 
(which is presumably fixed by a modulus field also)
and $R^{-1}$ is weaker since we
only require $R \stackrel{>}{\sim} 5 \; M^{-1}_{mess}$. 
In contrast, in the GUT model with 
$R \sim 3 M^{-1}$ there is only a ``modest'' hierarchy between
$R^{-1}$ and the $5D$ Planck scale $M$ (which, as mentioned
earlier, is assumed to
be {\em fundamental}).
} 
In any case, in the model
with $R \sim 5 \; M ^{-1} _ {GUT}$, 
the SUGRA and GM contributions to gaugino
masses will be (roughly)
same as in the model with 
$R \sim
3 M^{-1}$, i.e., $M_{1,2,3}$ are free parameters
which will generate positive sfermion (mass)$^2$ in RG scaling to the 
weak scale -- the only difference
is that in the model with $R \sim
3 M^{-1}$, there is a (negative) GM contribution to scalar (mass)$^2$ at 
$M_{GUT}$. The phenomeonlogy of these two models
should be similar (see section \ref{pheno}).

It is also clear from the 
discussion in section
\ref{problem} that, in general,
in models with (dominant) SUSY breaking in a GUT 
non-singlet (denoted by $\Sigma$), 
there
is a contribution 
to MSSM
gaugino masses
at one-loop (and to scalar (mass)$^2$ at two-loop)
from the coupling to gauge messengers -- to repeat, these are the
heavy gauge multiplets (with mass $\sim M_{GUT}$) which have a 
non-supersymmetric spectrum since the SUSY breaking field 
($\Sigma$) transforms
under the GUT gauge group. 
The (contributions to) MSSM gaugino masses generated at one-loop by
integrating out gauge messengers (at the scale
$M_{GUT}$) are generically (i.e., barring accidental
cancellations) given by
$\sim \alpha / \pi \; F_{\Sigma} \; M_{\not \! R} / M_{GUT}^2$, where 
$M_{\not \! R}$ is the $R$-symmetry breaking scale
-- thus the size of this contribution
is model-{\em dependent} due to $M_{\not \! R}$. The point is that
if the field $\Sigma$ also breaks the GUT symmetry (down to the SM 
gauge group), i.e., if the vev of the scalar component of $\Sigma$
($v_{\Sigma}$) is
$O(M_{GUT})$ (as in our model), 
then $M_{\not \! R} \sim M_{GUT}$. Therefore, in
a model with $F_{\Sigma} \neq 0$, {\em if}  
$v_{\Sigma}$ (or, in general, $M_{\not \! R}$) is $O(M_{GUT})$,
then
this gauge messenger
contribution to MSSM gaugino
masses is {\em comparable} to (and {\em independent} of) the
SUGRA contribution (from the operator
$\int d^2 \theta \; \Sigma / M_{Pl}
\; W_{\alpha} W^{\alpha} + h.c.$) \footnote{We assume that
$\Sigma$ is in a representation which appears in the symmetric
product of two adjoints.}
$\sim F_{\Sigma} / M_{Pl}$ 
(and is also
non-universal, in general). \footnote{There
might also be other GM contribution to gaugino (and scalar)
masses from, say,
``matter'' messengers (as in our model).}
Thus, as in our $SU(6)$ GUT model, the MSSM
gaugino mass relations are modified from that expected with only SUGRA
contribution -- in fact the model becomes {\em less} predictive (as far as
gaugino masses are concerned) since
there is an extra parameter corresponding to the gauge messenger
contribution.

There have been some 
recent studies of MSSM gaugino masses in
a scenario
with SUSY breaking by a GUT non-singlet so that SUGRA contribution
to gaugino masses is non-universal
\cite{anderson} --
in these studies, the above gauge messenger
contribution has {\em not} been mentioned. 
In the first reference in \cite{anderson}, it is 
{\em assumed}
that
$v_{\Sigma} \sim 0$, i.e., the SUSY breaking field has a small
vev in its scalar component. In this case, it is possible that
$M_{\not \! R} \ll M_{GUT}$ so that the gauge messenger contribution
to MSSM gaugino masses is small compared to the SUGRA contribution. 
Nonetheless, 
(in general) in order to be sure that the gauge messenger
contribution to MSSM gaugino masses is smaller than the SUGRA
contribution, the complete model has to be analysed to check that
$v_{\Sigma} \; (\hbox{or} \; M_{\not \! R}) \ll M_{GUT}$.
Also, a contribution to MSSM scalar (mass)$^2
\sim \left[ \alpha / \pi \right]^2 \left[ F_{\Sigma}
/M_{GUT} \right] ^2$ is generated at two-loops by
integrating out gauge messengers -- there is {\em no} suppression
due to $M_{\not \! R}$ unlike in the case of MSSM gaugino masses
(since scalar (mass)$^2$ do not break $R$-symmetry). Thus,
the gauge messenger contribution to scalar (mass)$^2$ is comparable to
(and again, independent of)
the SUGRA contribution to scalar (mass)$^2 \sim \left[ F_{\Sigma}/
M_{Pl} \right]^2$ (due to contact K\"ahler terms) 
{\em even if} $v_{\Sigma} \sim 0$
(i.e., the size of this contribution is model-{\em in}dependent). 
Furthermore, the gauge 
messenger contribution depends on gauge quantum numbers
and thus it is different for
squarks and sleptons (as in our $SU(6)$ GUT model), 
although it is flavor-conserving.

\section{Sparticle spectrum}
\label{pheno}
We now present briefly a sample sparticle spectrum in the $5D$
model. 
\subsection{Parameters of the model}
In this model, the MSSM sfermion masses (at $M_{GUT}$)
and gaugino masses 
are determined by three parameters --
$F_v/v \; (v \sim M_{GUT})$, $c_2$ and $c_3$. 
The GM contribution (Eqs. (\ref{gmgaugino}) and (\ref{gmscalar}))
can be written in terms of $F_v/v$ (assuming that the beta-functions
\footnote{As mentioned earlier, even though the $N \bar{P} \bar{H}$
coupling is not 
allowed (at the renormalizable level) in the $5D$ model, 
we assume that the additional $({\bf 5} + \bar{{\bf 5}})$
in $N$, $\bar{P}$'s have a mass of $M_{GUT}$ and so the
values of the beta-functions are the same
as in section \ref{review} (the $4D$ model), 
i.e., $b_6 = -11$ and $b_A$'s are the 
usual MSSM
beta-functions.} and
gauge couplings 
are fixed)
and the SUGRA contrbution
to MSSM gaugino masses, Eq. (\ref{sugragaugino}),
depends on $F_v/v$ and $c_{2,3}$ (any uncertainty in the ratio
of $v \sim M_{GUT}$ and $M$ (or $M_{Pl}$) can be absorbed
into $c_{2,3}$).
For convenience, we choose the gaugino
masses at the GUT scale, $M_1$, $M_2$ and $M_3$ (which 
are combinations of these parameters) to be the 
free
parameters. \footnote{Note that if only the adjoint field (and not singlets)
breaks SUSY, then $c_3 = 0$ in Eq.
(\ref{sugragaugino}) and one has only
two free parameters.}
Using Eqs. (\ref{gmgaugino}) and (\ref{sugragaugino}), we get:
\begin{equation}
\frac{1}{36} \; ( - 5 M_1 + 3 M_2 + 2 M_3) = \frac{\alpha _6}{4 \pi}
\; \frac{F_v}{v}
\label{Fv}
\end{equation}
which parametrizes the GM contribution.
This relation is used to determine
the GM contribution to scalar (mass)$^2$ at the GUT scale in
terms of $M_A$'s using Eq. (\ref{gmscalar})
with $\mu \approx M_{GUT}$. 
The scalar masses are then evolved to the weak scale
using the one-loop RG equations.

We note that the SUSY-GUT prediction for 
$\sin ^2 \theta _w$ (in terms of $\alpha _s$ and $\alpha _{em}$)
is affected by the operator
$W_{\alpha} W^{\alpha} \Sigma / M^2$ in Eq. (\ref{sugragaugino}) 
\cite{hall} -- in this model, this effect (which is at the
$M_{GUT} / M_{Pl}$, i.e., percent level) is 
related to SUGRA contribution to gaugino mass.
The method used to give mass to extra
doublets in $H$, $\bar{H}$ (see section \ref{review}) also
affects the prediction for $\sin ^2 \theta _w$ --
in this case this pair of doublets gets a mass of $O( M ^2 _{GUT}
/ M ) < M_{GUT}$ (from the superpotential
$W_4$) which shifts the $\sin ^2 \theta _w$
prediction by about a percent.
Since we wish to illustrate the main
features of the spectrum in this paper, we
will neglect these effects (which might cancel each other).

We also
neglect RG scaling of sfermion masses (at one-loop due to SUGRA contribution
to gaugino
masses and at two-loops due to SUGRA contribution ($\sim F_v / M_{Pl}$) to
$Q$, $\bar{Q}$ etc. masses) 
between the ($5D$) Planck scale $M$ and the 
GUT scale (sfermion masses
are negligible at the scale $M$) \footnote{This
contribution is positive and
flavor-conserving -- thus it makes sfermions with same gauge
quantum numbers
more degenerate
and in particular the RH slepton heavier (see later).}
-- the RG logarithm $\sim
\ln \left( M / 
M_{GUT} \right)$ is much smaller than that for
RG scaling between GUT and weak scale (of course, the larger group
theory factors above $M_{GUT}$ might compensate for the smaller
RG logarithm as emphasized in the context of
extra dimensional models in \cite{skiba}). \footnote{A detailed study
of the phenomenology
in this GUT model (including the effects
of RG scaling between $M$ and $M_{GUT}$)
is in progress.} The (SUGRA mediated) gaugino masses also run between $M$ 
and
$M_{GUT}$; however since $M_A / \alpha _A$ is RG-invariant (at
one-loop), the ratio of the MSSM gaugino masses remains the same
in this RG scaling since the gauge couplings are unified (of course,
at $M_{GUT}$, the MSSM gauginos get additional contribution
to their masses (Eq. 
(\ref{gmgaugino})).

Since the usual Higgs doublets propagate in the extra dimension, soft
SUSY breaking Higgs (mass)$^2$ and also $B \mu$ and $\mu$ 
(Giudice-Masiero mechanism \cite{gm})
are generated by the SUGRA interactions:
\begin{eqnarray}
{\cal L} & \sim & \int d^4 \theta \left( \frac{1}{M^3}
h^{\dagger} h \left[\Sigma ^{\dagger} \Sigma + ..\right]
+ \frac{1}{M^3}
\bar{h} \bar{h} ^{\dagger} \left[ X^{\dagger} X + ..\right]
\right) + \nonumber \\ 
 & & \int d^4 \theta
\left( h \bar{h} \frac{1}{M^2} \left[ \Sigma ^{\dagger} + X
^{\dagger} + \bar{X}
^{\dagger}
\right] 
+ h \bar{h} \frac{1}{M^3} \left[ \Sigma ^{\dagger} \Sigma
+ ..\right] + h.c. \right),
\end{eqnarray}
where $h$, $\bar{h}$ are $5D$ fields. The couplings of zero-modes of $h$,
$\bar{h}$ (which correspond to the light $4D$ Higgs fields) 
are suppressed by
a factor of $\sqrt{R}$ (to account for the normalization of the zero mode)
so that 
$m _{H_{u,d}, SUGRA} ^2 \sim F_v ^2 / M^3 \; 1/ R$,
$\mu \sim F_v / M^2 \; 1/ R$ and $B \mu \sim F_v ^2 / M^3 \; 1/ R$. 
Since $MR \sim 3$ and $M_{Pl}^2 \sim M ^3 R$, we see that
$m _{H_{u,d}, SUGRA} ^2$, 
$\mu ^2$, $B\mu$
are all of order $\left( F_v / M_{Pl} \right)^2$, i.e, 
of the same order as, but independent of, gaugino
and sfermion masses (they are also
independent of each other).
Of course, the Higgs doublets also get
soft (mass)$^2$ from GM (Eq. (\ref{gmscalar}), which are related
to the other sfermion and gaugino masses. We choose
the Higgs soft (mass)$^2$ at the GUT scale (which are the sum of the 
SUGRA and GM contributions) to be free parameters, denoted
by $m_{H_u}^2$ and $m_{H_d}^2$.

In this model, the gauginos of the heavy gauge multiplet have
a SUSY breaking mass $\sim F_v / M_{GUT}$ since the SUSY breaking field
is an adjoint under the GUT gauge group.
This generates trilinear 
(MSSM) scalar terms of $O( \alpha / \pi F_v / M_{GUT} )
\sim F_v / M_{Pl}$ at the GUT scale (when the heavy gauginos are integrated
out). The exact expression is \cite{gr}:
\begin{eqnarray}
V  & \ni  & \sum _i A_i Q _i \partial _{Q _i} W (Q), \\ \nonumber
A_i (\mu_{RG}) 
& = & \frac{ \partial \ln Z _{Q_i} (v ^{\dagger}, v, \mu_{RG}) }
{\partial \ln v} \; \frac{F_v}{v}. 
\end{eqnarray}
In this case, we have
\begin{eqnarray}
A_i (M_{GUT}) & = & \frac{F_v}{v} \frac{\alpha _6}{4 \pi}
\left( 2 C^i_6 - 2 \sum _A C^i_A \right).
\end{eqnarray}
We neglect all Yukawa couplings other than the top quark coupling and so
only the coupling
$\lambda _t A_t H_u \tilde{Q} _3 \tilde{t}^c$ is non-zero and is given by
\begin{eqnarray}
A_t (M_{GUT}) & = & 15.3 \frac{F_v}{v} \frac{\alpha _6}{4 \pi}.  
\end{eqnarray}
Thus, the $A$-term at the GUT scale depends only on $F_v/v$ 
(and gauge couplings), i.e., it is {\em not}
an independent parameter -- in particular, there is no
SUGRA contribution to $A_t$ since the top squark and the SUSY
breaking fields are localized on separate branes. \footnote{As mentioned
earlier, we neglect RG scaling between GUT and Planck scales; this effect
does generate (due to non-vanishing gaugino masses)
a small $A_t$ term at the GUT scale which depends on
the SUGRA contribution to gaugino masses, i.e., $F_v/v$, $c_2$ and $c_3$.}

Thus, the {\em fundamental} parameters in this model are:
$F_v/v$, $c_{2,3}$, $m_{H_{u,d}}^2$, $B\mu$, $\mu$ and $\lambda _t$
(top quark Yukawa coupling). Two of these parameters are fixed by
the observed values of $m_Z$ and $m_t$
so that the {\em free} (i.e., input) parameters can be chosen to be
$M_{1,2,3}$ (which, as explained earlier are combinations
of $F_v/v$ and $c_{2,3}$), $m_{H_{u,d}}^2$ and $\tan \beta$;
$\mu$ and $B \mu$ can then be determined in terms of these
parameters as usual using the minimization conditions for the Higgs
potential.

\subsection{Sample sparticle spectrum}
\renewcommand{\arraystretch}{1}
\begin{table}
\begin{center}
\begin{tabular}{|c|c|}
\hline
$M_{\chi ^0 _1}$ & $105$, $120$ \\ \hline
$M_{\chi ^0 _2}$ & $165$ \\ \hline
$M_{\chi ^0 _3}$ & $190$ \\ \hline
$M_{\chi ^0 _4}$ & $290$ \\ \hline
$M_{\chi ^+ _1}$ & $145$, $170$ \\ \hline
$M_{\chi ^+ _2}$ & $290$ \\ \hline
$m_{\tilde{e}_R}$ & $115$ \\ \hline
$m_{\tilde{e}_L, \tilde{\nu}_L}$ & $210$, $200$ \\ \hline
$M_{\tilde{g}}$ & $365$ \\ \hline
$m_{\tilde{u}_R}$ & $315$ \\ \hline
$m_{\tilde{d}_R}$ & $320$ \\ \hline
$m_{\tilde{u}_L, \tilde{d}_L}$ & $375$, $380$ \\ \hline
$m_{\tilde{t} _{1,2}}$ & $190$, $400$ \\
\hline
\end{tabular}
\end{center}
\caption{Sample spectrum in the model for 
renormalization scale
$\mu _{RG} \approx 500$ GeV and 
the input
parameters: $M_1 = M_2 = -300$, $M_3 = -150$, $m_{H_u}^2 = (150)^2$,
$m_{H_d}^2 = (300)^2$ and $\tan \beta = 5$ which give $\mu =
180$ (all masses in GeV). 
The two values for $M_{\chi ^0 _1}$ and $M_{\chi ^+ _1}$ are for different
signs of $\mu$ (the other neutralino/chargino and stop
masses do not depend strongly on the sign of $\mu$).}
\label{spectrum}
\end{table}

In Table \ref{spectrum}, a sample spectrum is
presented for the input
parameters $M_1 = M_2 = -300$ GeV and
$M_3 = -150$ GeV,
$m_{H_u}^2 = (150 \; \hbox{GeV})^2$, $m_{H_d}
^2 = (300 \; \hbox{GeV})^2$ and
$\tan \beta = 5$. 
We have included 
the electroweak $D$-term and Fayet-Illiopoulos (FI)
$D$-term contributions to scalar (mass)$^2$ which are given by
$m_Z^2 \cos 2 \beta \left( T_3 - Q \sin ^2 \theta _w \right)$ and
$ - 0.053 \; Y \left( m_{H_u}^2 - m_{H_d}^2 \right)$ 
\footnote{The
FI $D$-term contribution in $\tilde{g}$MSB was emphasized in
\cite{gmsb2}.}, respectively. \footnote{We assume for simplicity
that there is no $D$-term contribution from the breaking of the extra
$U(1)$ (of $SU(6)$) at $M_{GUT}$.}
The mixing between the top squarks and the one-loop
corrections to the effective
Higgs potential (due to top
quark and top squark masses only)
have been included.

Some of the characteristic features of the spectrum are as follows.

Gaugino masses are non-universal in general
since both the GM contribution and a part of the
SUGRA contribution (due to the operator
$\int d^2 \theta \; c_2 \; \Sigma / M^2 \;
W_{\alpha} W^{\alpha}$) are non-universal
(the relative GM contributions to gaugino masses (Eq. (\ref{gmgaugino})) are 
model-dependent due to dependence on $b_6$).
Models with non-universal gaugino masses 
(at the GUT/Planck scale) have been studied earlier
\cite{others}. 
In most of these models,
sfermion masses are independent 
parameters
whereas in our GUT model
the sfermion masses are determined in terms of 
the gaugino masses ($M_{1,2,3}$).
\footnote{In
``$D$-brane'' models \cite{dbrane}, it is possible that $M_2 \neq M_3$
if $SU(3) _c$ and $SU(2) _w$ originate from different
$D$-brane sectors. Since quark doublets transform under
all three SM gauge groups,
$U(1)_Y$ has to orginate in either of these two sectors -- this implies
that $M_1 = M_2$ or $M_1 = M_3$, unlike the GUT model where it is possible
that {\em all three} gaugino masses are different.}

No-scale SUGRA models have vanishing
scalar masses at $M_{Pl}$ and gaugino masses (which can be
non-universal) as usual drive sfermion
(mass)$^2$ positive in RG scaling to the weak scale
($\tilde{g}$MSB with non-universal gaugino masses
has similar boundary conditions).
However, in the GUT
model
studied here, there is a GM contribution to scalar masses at the GUT scale
and also there are SUGRA contributions to Higgs soft masses (which, in turn,
result in a FI $D$-term contribution to sfermion masses at the weak scale).
Thus, the GUT model can (in principle) be distinguished
from no-scale SUGRA models with non-universal gaugino masses
by precision sparticle spectroscopy.

The (GM contribution to) RH slepton (mass)$^2$ at the GUT scale 
is negative (Eq. (\ref{gmscalar})) 
and therefore RH slepton 
(its (mass)$^2$
is driven positive by bino mass) 
and $\chi _1 ^0$ (which is roughly
the bino) are close in mass. \footnote{In this sample spectrum,
the 
FI $D$-term (positive for RH slepton)
makes the RH slepton (slightly) heavier than $\chi _1 ^0$
(for one sign of $\mu$).}
For the same reason, 
the mass splitiing between the left-handed slepton and 
the RH slepton is large \footnote{If the GM contribution
at $M_{GUT}$ is neglected, then we get $m_{\tilde{e}_R} \approx 140$
GeV while $m_{\tilde{e}_L}$ remains about the same.}
even though we have chosen
$M_1 = M_2$ for the above sample spectrum,
i.e., usually we expect
$m_{\tilde{e}_L} 
- m_{\tilde{e}_R}$ to be large only if $M_2 > M_1$
(due to gaugino mass contributions in RG scaling).
As mentioned earlier, if we include RG scaling 
between $M$ and
$M_{GUT}$, then {\em all} sfermions
will be heavier leading to a {\em larger}
mass splitting between RH slepton and $\chi ^0 _1$ (which will be the
lightest supersymmetric particle (LSP)) whereas the large mass splitting
between $\tilde{e}_L$ and $\tilde{e}_R$ (which is
due to the negative GM contribution
to $m_{\tilde{e}_R}^2$ at $M_{GUT}$) will remain about the {\em same}.

The lower limit on the RH slepton mass $\sim 90$ GeV fixes (roughly)
a
minimum value for $M_1$. But since the three gaugino masses
are independent parameters, $M_3$ can be smaller than $M_{1,2}$
so that there is not much of a hierarchy (in masses)
between squarks/gluino and sleptons as seen in Table \ref{spectrum}.
Also, 
since $M_3$ can be smaller than $M_{1,2}$
and also (GM mediated)
stop (and other
squark) (mass)$^2$ are small and negative at the GUT scale, $|m_{H_u}^2|$
at the weak scale (as usual the up-type Higgs (mass)$^2$
is driven negative by stop (mass)$^2$ and gluino mass)
and hence $\mu$ can be small (in this case $\sim 180$ GeV),
thus reducing the fine-tuning in electroweak symmetry breaking -- 
small $\mu$ also results in ``light'' chargino/neutralino. \footnote{As 
mentioned earlier, RG scaling between $M_{GUT}$ and $M$ will
give a positive contribution to scalar (mass)$^2$ due to
(SUGRA mediated) gaugino masses -- this contribution is about 
the same for all scalars (squarks, sleptons
and Higgs)
because of the unified gauge symmetry, unlike the case of
RG scaling {\em below} the GUT scale where, since
$\alpha _3 > \alpha _{1,2}$ the gluino mass contribution
(to squark masses) is larger (assuming universal gaugino masses). Thus,
the above features, i.e., 
the ``small'' hierarchy between squarks and sleptons and small $\mu$
will {\em not} be affected by the inclusion of this effect.} 

This should be compared to $\tilde{g}$MSB with
{\em universal} gaugino mass \cite{gmsb1, gmsb2}
where a minimum value for $M_1$ 
fixed by RH slepton mass implies a minimum value
for $M_3$ 
(typically $\stackrel{>}{\sim} 200$ GeV)
which results in a larger hierarchy betwen slepton
and squark/gluino masses and also
larger $|m_{H_u}^2|$ and hence fine tuning (due to larger $\mu$).
In a minimal gauge mediation model also, there is a large hierarchy
between squark/gluino masses and slepton masses (since masses are
proportional to gauge couplings). In minimal SUGRA mediated SUSY
breaking (with universal scalar mass, $m_0$, and universal gaugino mass, 
$M_{1/2}$) it is possible to have small hierarchy between sleptons
and squarks/gluino. In any case, non-universal gaugino masses
distinguishes these models from our model.  

Of course, we
expect that with extra parameters as
compared to a minimal model (especially non-universal gaugino masses)
such a spectrum can be attained. However in this model
these extra parameters are {\em not ad hoc}, but are
{\em well-motivated} -- they are 
justified by the way SUSY is broken in the model.

\section{Summary}
A model in which the 
{\em same} scalar
potential breaks SUSY and a 
GUT symmetry was presented in
\cite{gutsusy} -- this model has {\em dynamical} origins
for {\em both} SUSY breaking and GUT scales.
In this model, the
SUGRA and gauge mediated contributions
to scalar and gaugino masses are comparable -- this
enables a viable spectrum to be attained since the gauge
mediated contribution to RH slepton (mass)$^2$ by itself is
negative. But, the flip side is that the SUGRA contribution
to sfermion masses (from non-renormalizable contact
K\"ahler terms) results in flavor violation. 

In this paper,
we suggested that this
``problem'' will be present in {\em any} model
in which the same field breaks SUSY and a GUT symmetry and 
demonstrated that, 
using an extra spatial dimension,
positive and (at the same time)
flavor-conserving
sfermion (mass)$^2$ can be obtained in this model.
The model has {\em non}-universal gaugino masses and sfermion masses
are predicted in terms of gaugino masses. The hierarchy between squark/gluino
masses and slepton masses can be small and (typically) a large mass
splitting between RH and LH slepton is expected. 
 
{\bf Acknowledgments} 

The author thanks Markus Luty and Nir
Polonsky for suggestions and Neal Weiner for comments.
This work is supported by DOE Grant DE-FG03-96ER40969.

\end{document}